# Gap Solitons in a Nonlinear Quadratic Negative Index Cavity

Michael Scalora[1], Domenico de Ceglia[1,2], Giuseppe D'Aguanno[1], Nadia Mattiucci[3,1],

Neset Akozbek[3], Marco Centini[4], Mark J. Bloemer[1]

1   Charles M. Bowden Research Center, AMSRD-AMR-WS-ST,

Research, Development, and Engineering Center, Redstone Arsenal, AL 35898-50003

2   Dipartimento di Elettrotecnica ed Elettronica, Politecnico di Bari, Via Orabona 4,

70124 Bari, Italy

3   Time Domain Corporation, Cummings Research Park 7057 Old Madison Pike

Huntsville, Alabama 35806, USA

4   INFM at Dipartimento di Energetica, Universita di Roma 'La Sapienza', Via A.

Scarpa 16, 00161 Roma, Italy

**ABSTRACT**

By integrating the full Maxwell's equations we predict the existence of gap solitons in a quadratic, Fabry-Perot negative index cavity. An intense, fundamental pump pulse shifts the band structure that forms when magnetic and electric plasma frequencies are different, so that a weak, second harmonic pulse initially tuned inside the gap is almost entirely transmitted. The process is due cascading, which occurs far from phase matching conditions, and causes pulse compression. A nonlinear polarization spawns a dark soliton, while a nonlinear magnetization produces a bright soliton.

The term "gap soliton" was coined to describe the shape that the electric field assumes when an incident, continuous wave beam is tuned inside the photonic band gap of a one-dimensional, periodic structure of finite length, so that a third order ($\chi^{(3)}$) nonlinearity causes the beam to be transmitted [1]. The physics of how such a state may be excited is exceptionally simple: a nonlinear change in the intensity-dependent refractive index of at least one of the constituent materials causes a shift of the photonic band edge, thus placing the incident beam within the pass band, and allowing its transmission. An excellent review of third order gap solitons may be found in reference [2]. Transverse, diffractive spatial gap solitons have been predicted in quadratic ($\chi^{(2)}$) materials in the context of multilayer structures [3], and are generally due to cascading, a process that occurs when pump and second harmonic beams interact far from phase matching conditions. Temporal, two-color gap solitons were also predicted in quadratic, shallow-depth Bragg gratings [4], and typically rely on doubly resonant conditions, and strong coupling between the fundamental (FF) and second harmonic (SH) beams. The recent interest in negative index materials (NIMs) [5] has led to predictions of $\chi^{(3)}$ gap solitons [6] near the band edge of the intrinsic gap of a Fabry-Perot, NIM cavity [7] in the form of a single slab of material immersed in vacuum. Unlike the zero average-index gap [8], formation of the intrinsic band structure does *not* require the presence of a positive index material (PIM), and it is a peculiarity of the frequency range where a NIM has dielectric susceptibility and magnetic permeability of opposite signs [7]. The peculiarities of the band structure extend to the field localization properties, which appear to be unique even for a single slab of material [6, 7].

In this Letter we report second harmonic gap solitons in a $\chi^{(2)}$-active NIM etalon. For a positive nonlinear coefficient, one may excite either a dark or a bright soliton, depending on whether an electric or magnetic nonlinearity is present. As is well-known in the case of ordinary PIMs, the FF and SH fields do not exchange energy if the relative phase difference between the incident fields is chosen so that the interaction proceeds far from the phase matching condition, thus triggering cascading, and the interaction resembles a $\chi^{(3)}$ process. In a NIM etalon, the formation of a gap soliton at the second harmonic frequency follows a similar pattern, with some distinguishing characteristics. An intense FF pulse is tuned to the first resonance, on the low frequency side of the intrinsic band gap, where the index of refraction is negative, and it is mostly transmitted. Although we choose the FF to be resonant, it is not a necessary condition but it helps to lower the nonlinear thresholds. A much weaker SH pulse is then tuned inside the gap, where the index of refraction is near zero (n~$10^{-3}$), in proximity of the high frequency band edge, so that in the absence of nonlinear coupling it is mostly reflected. Then, by properly adjusting the relative input phase difference between the fields, the SH pulse experiences no net gain, causing a dynamic shift of the band edge. As a result, the SH pulse is effectively pushed out of the band gap, it is spatially and temporally compressed, and almost completely transmitted: transmittance switches from 4% to about 90%.

To model the dynamics of interacting, short pulses in a NIM cavity we begin by writing nonlinear, second order polarization and magnetization as $\mathbf{P}_{NL} = \chi_P^{(2)} \mathbf{E} \bullet \mathbf{E}$, and $\mathbf{M}_{NL} = \chi_M^{(2)} \mathbf{H} \bullet \mathbf{H}$, where $\chi_P^{(2)}$ and $\chi_M^{(2)}$ are the respective electric and magnetic nonlinear coefficients. We assume linearly polarized fields of the type:

$$\begin{aligned}\mathbf{E} &= \hat{\mathbf{x}}\left(\mathcal{E}_\omega(z,t)e^{i(kz-\omega_0 t)} + c.c + \mathcal{E}_{2\omega}(z,t)e^{2i(kz-\omega_0 t)} + c.c\right) \\ \mathbf{H} &= \hat{\mathbf{y}}\left(\mathcal{H}_\omega(z,t)e^{i(kz-\omega_0 t)} + c.c + \mathcal{H}_{2\omega}(z,t)e^{2i(kz-\omega_0 t)} + c.c\right)\end{aligned} \quad , \qquad (1)$$

where $k = \omega_0/c$ is the free space wave vector, $\omega_0$ is the corresponding carrier frequency, $\mathcal{E}_{\omega,2\omega}(z,t)$ and $\mathcal{H}_{\omega,2\omega}(z,t)$ are general, complex envelope functions about which no approximations are made. The fields are assumed to be initially located in free space, to reflect the choice of initial wave vector. The nonlinear polarization and magnetization may then also be described in terms of generic envelope functions, and carrier wave vector and frequency as follows:

$$\begin{aligned}\mathbf{P}_{NL} &= \hat{\mathbf{x}}\left(\mathcal{P}_\omega(z,t)e^{i(kz-\omega_0 t)} + c.c + \mathcal{P}_{2\omega}(z,t)e^{2i(kz-\omega_0 t)} + c.c\right) \\ \mathbf{M}_{NL} &= \hat{\mathbf{y}}\left(\mathcal{M}_\omega(z,t)e^{i(kz-\omega_0 t)} + c.c + \mathcal{M}_{2\omega}(z,t)e^{2i(kz-\omega_0 t)} + c.c\right)\end{aligned} \quad , \quad (2)$$

where $\mathcal{P}_\omega(z,t) = 2\chi_P^{(2)}\mathcal{E}_\omega^*\mathcal{E}_{2\omega}$, $\mathcal{P}_{2\omega}(z,t) = \chi_P^{(2)}\mathcal{E}_\omega^2$, $\mathcal{M}_\omega(z,t) = 2\chi_M^{(2)}\mathcal{H}_\omega^*\mathcal{H}_{2\omega}$, and $\mathcal{M}_{2\omega}(z,t) = \chi_M^{(2)}\mathcal{H}_\omega^2$. The inclusion of linear dispersion is straight forward, and in the limit where second and higher order dispersion may be neglected, Maxwell's equations take the following form [9, 10]:

$$\begin{aligned}\alpha_{\tilde{\omega}}\frac{\partial \mathcal{E}_{\tilde{\omega}}}{\partial \tau} &= i\beta\left(\varepsilon_{\tilde{\omega},\xi}\mathcal{E}_{\tilde{\omega}} - \mathcal{H}_{\tilde{\omega}}\right) - \frac{\partial \mathcal{H}_{\tilde{\omega}}}{\partial \xi} + 4\pi\left(i\beta\mathcal{P}_{\tilde{\omega}} - \frac{\partial \mathcal{P}_{\tilde{\omega}}}{\partial \tau}\right) \\ \gamma_{\tilde{\omega}}\frac{\partial H_{\tilde{\omega}}}{\partial \tau} &= i\beta\left(\mu_{\tilde{\omega},\xi}\mathcal{H}_{\tilde{\omega}} - \mathcal{E}_{\tilde{\omega}}\right) - \frac{\partial \mathcal{E}_{\tilde{\omega}}}{\partial \xi} + 4\pi\left(i\beta\mathcal{M}_{\tilde{\omega}} - \frac{\partial \mathcal{M}_{\tilde{\omega}}}{\partial \tau}\right) \\ \alpha_{2\tilde{\omega}}\frac{\partial \mathcal{E}_{2\tilde{\omega}}}{\partial \tau} &= 2i\beta\left(\varepsilon_{2\tilde{\omega},\xi}\mathcal{E}_{2\tilde{\omega}} - \mathcal{H}_{2\tilde{\omega}}\right) - \frac{\partial \mathcal{H}_{2\tilde{\omega}}}{\partial \xi} + 4\pi\left(i2\beta\mathcal{P}_{2\tilde{\omega}} - \frac{\partial \mathcal{P}_{2\tilde{\omega}}}{\partial \tau}\right) \\ \gamma_{2\tilde{\omega}}\frac{\partial H_{2\tilde{\omega}}}{\partial \tau} &= 2i\beta\left(\mu_{2\tilde{\omega},\xi}\mathcal{H}_{2\tilde{\omega}} - \mathcal{E}_{2\tilde{\omega}}\right) - \frac{\partial \mathcal{E}_{2\tilde{\omega}}}{\partial \xi} + 4\pi\left(i2\beta\mathcal{M}_{2\tilde{\omega}} - \frac{\partial \mathcal{M}_{2\tilde{\omega}}}{\partial \tau}\right)\end{aligned} \quad , \quad (3)$$

where $\alpha_{\tilde{\omega}} = \frac{\partial[\tilde{\omega}\varepsilon(\xi)]}{\partial\tilde{\omega}}|_{\omega_0}$, $\gamma_{\tilde{\omega}} = \frac{\partial[\tilde{\omega}\mu(\xi)]}{\partial\tilde{\omega}}|_{\omega_0}$, and both ε and μ are complex functions of frequency and of the spatial coordinate. We have chosen $\lambda_0 = 1\mu m$ as the reference wavelength, and have adopted the following scaling: $\xi = z/\lambda_0$ is the scaled longitudinal coordinate; $\tau = ct/\lambda_0$ is the time in units of the optical cycle; $\beta = 2\pi\tilde{\omega}$ is the scaled wave vector; $\tilde{\omega} = \omega/\omega_0$ is the scaled frequency, and $\omega_0 = 2\pi/\lambda_0$.

Eqs.(3) contain no approximations other than the assumption that the medium is isotropic, and that higher order material dispersion terms may be neglected. The latter point can easily be justified, since we are considering a material only two wavelengths thick. We note that in NIMs typical higher order dispersion lengths range from a few tens [11] to a few thousand wavelengths, and that the neglect of higher order, material dispersion terms does not necessarily restrict the envelope functions to be slowly varying in time. Eqs.(3) thus provide a very accurate physical picture of the dynamics even for pulses down to just a few wave cycles in duration [9, 10]. These observations are fully confirmed by integrating the full Maxwell equations coupled to a set of driven, nonlinear oscillators, which we perform using a finite difference, time domain integration technique, yielding identical results.

The intrinsic band gap, the relative tuning of the FF and SH fields, and the index of refraction are depicted in Fig.(1), given a Drude model with the following characteristics: $\varepsilon(\tilde{\omega}) = 1 - \frac{\tilde{\omega}_E^2}{\tilde{\omega}^2 + i\tilde{\gamma}\tilde{\omega}}$, $\mu(\tilde{\omega}) = 1 - \frac{\tilde{\omega}_M^2}{\tilde{\omega}^2 + i\tilde{\gamma}\tilde{\omega}}$, where $\tilde{\omega}_E = \omega_E/\omega_0$ and $\tilde{\omega}_M = \omega_M/\omega_0$ are the scaled electric and magnetic plasma frequencies, respectively, and $\tilde{\gamma}$ is the scaled damping coefficient. As a representative example, we choose $\tilde{\omega}_M = 1$, $\tilde{\omega}_E = 0.5205$, $\tilde{\gamma} = 10^{-4}$. Therefore, the *magnetic* plasma frequency coincides with our reference wavelength. Although it is never explicitly invoked in the integrations of Maxwell's equations, the index of refraction may be retrieved as usual, i.e.

$n(\tilde{\omega}) = \pm\sqrt{\varepsilon(\tilde{\omega})\mu(\tilde{\omega})}$, and the negative root is chosen when both $\varepsilon$ and $\mu$ are simultaneously negative [5]. Our choice of $\tilde{\gamma}$ corresponds to an absorption length of approximately 500 microns for the pump, and ten times larger for the SH pulse, so that neither are appreciably attenuated, as the transmittance curve of Fig.(1) shows. However, our calculations show that the soliton does not loose its coherence with the introduction of more significant absorption, whose presence, in fact, tends to simply raise the nonlinear thresholds. For example, taking $\tilde{\gamma} = 10^{-3}$, which corresponds to an absorption length of ~50μm, causes the transmittance of the FF to drop to ~77%. Our calculations show that this loss may be compensated by a ~30% increase in the peak intensity of the FF pulse.

Unlike the damping coefficient, $\tilde{\omega}_M$ and $\tilde{\omega}_E$ contain more subtleties, primarily because their relative magnitudes determine how the fields will become localized [7]. We note that experimentally, $\tilde{\omega}_M$ and $\tilde{\omega}_E$ may be set by properly engineering the size of the elemental, split-ring resonator circuit [12], or by properly managing the geometry of the various components [13]. Then, our choice of smaller electric plasma frequency causes the electric field to become highly localized (a single maximum) at the low frequency band edge, and anti-localized (a single minimum) at the high frequency band edge, near the second harmonic frequency. Exchanging the values of $\tilde{\omega}_M$ and $\tilde{\omega}_E$ causes the electric and magnetic fields to trade roles, as an analysis of Eqs.(3) suggests. Consequently, the nonlinear polarization generates anti-localized, dark gap solitons, and a nonlinear magnetization induces localized, bright, gap solitons. In Fig.(2) we show the anti-localized and localized states that respectively correspond to the excitation of dark and bright solitons, for a SH frequency tuned near the high frequency band edge resonance. In Fig.(3) we depict the energy contained in the SH pulse as a function of the relative input phase difference between the two incident pulses, normalized in units of the incident, SH energy. When $\delta\varphi \sim 10.5^o$, the energy exchanged between the fields amounts to less than one part in a thousand. The temporal dynamics of the integrated SH energy that corresponds to $\delta\varphi = 10.5^o$ is shown in Fig.(4). Each point on the curves of Fig.(3) was obtained using incident pulses approximately 1ps in duration, or about 200 optical cycles; the intensity of the FF ($\sim 100 MW/cm^2$) is approximately $2 \times 10^5$ greater than the SH peak intensity, so that the FF field propagates undisturbed; and $\chi_P^{(2)} \sim 8 pm/V$, $\chi_M^{(2)} = 0$. In Fig.(5) we show incident and scattered SH pulse intensities, normalized to incident peak intensity. The figure reveals that in addition to being mostly transmitted, the pulse is compressed by ~30%. One should contrast these results with linear behavior, i.e. Fig.(1), where we obtain ~4% transmission. Both these aspects of the dynamics, i.e., high transmittance and pulse compression, along with the fields shown in Fig.(2), are fully consistent with soliton-like behavior, as observed experimentally in the case of a soliton that forms and propagates near the band edge of a fiber Bragg grating [14].

The dynamics that we have described depend on a number of factors. For instance, the curves of Fig.(3) are sensitive not only to initial relative phase difference, but also to peak pulse intensities. Because of this sensitivity, pulse duration is also important, as pulse bandwidth determines the degree of localization light achieves inside the etalon. Thus the kind of gap soliton that we report persists well into the sub-picosecond regime, and as we have seen even when absorption is present, albeit with relatively higher nonlinear thresholds.

In conclusion, we report SH bright and dark gap solitons in a 2-μm thick, nonlinear, quadratic, Fabry-Perot cavity. The relative phase difference between the incident fields is chosen to induce cascading processes and a dynamic shift of the intrinsic band edge that causes pulse compression and gap soliton formation. The soliton is relatively impervious to the introduction of absorption and the reduction of pulse width. These findings are relevant in the optical regime, as negative index materials are actually being fabricated in the near IR region [13], with good prospects for devices in the visible part of the spectrum. We thus hope that our results, which take into account effects of finite size and material absorption, will further stimulate research in this direction. Finally, we also note that gap solitons at the fundamental frequency may also be created using a negative nonlinear coefficient, and by reversing the roles and intensities of the SH and FF pulses.

**Acknowledgement**

G.D. thanks the National Research Council for financial support.

**Figure Captions**

**Fig.(1):** Transmittance (left axis) and real part of the index of refraction (right axis) for the Drude model described in the text. The FF pulse is tuned at $\omega=0.5$, which coincides with the first band edge resonance on the low frequency side. The SH pulse is tuned to $\omega=1$, which falls inside the gap.

**Fig.(2):** Dashed curve: dark SH soliton. This state is excited by the pump field interacting with a nonlinear polarization, as the band edge dynamically shifts and tunes the SH frequency near the first high frequency band edge resonance. The bright, SH soliton (thick, solid curve) is excited via a nonlinear magnetization. The thin, solid line traces the real part of the dielectric function.

**Fig.(3):** Integrated SH energy as a function of relative phase difference between incident FF and SH pulses, normalized with respect to incident SH energy. A total energy value near unity thus corresponds to a point of no net energy exchange between the field, and to maximum dynamics shift of the band edge.

**Fig.(4):** Temporal dynamics of the total, transmitted (to the right of the structure in Fig.(4)), and reflected (to the left of the structure in Fig.(4)) SH energies for $\delta\varphi \sim 10.5^o$. Transient behavior is evident in the total, instantaneous energy. The overall transmission settles to approximately 90%. The generic field envelopes are of the type: $\mathcal{E}_\omega, \mathcal{H}_\omega(\xi,0) = \mathcal{E}_0^{(\omega)}, \mathcal{H}_0^{(\omega)} e^{-(\xi-\xi_0)^2/w^2} e^{i\varphi}$ for the fundamental, and $\mathcal{E}_{2\omega}, \mathcal{H}_{2\omega}(\xi,0) = \mathcal{E}_0^{(2\omega)}, \mathcal{H}_0^{(2\omega)} e^{-(\xi-\xi_0)^2/w^2}$ for the second harmonic fields. In our units, the choice of $w \sim 200$ corresponds to a pulse with full width at half maximum of $\sim$1ps.

**Fig.(5):** Incident and scattered SH pulses. The figure shows that the peak intensity of the transmitted pulse increases by approximately 20%, and its full width at half maximum decreases by approximately 30%.

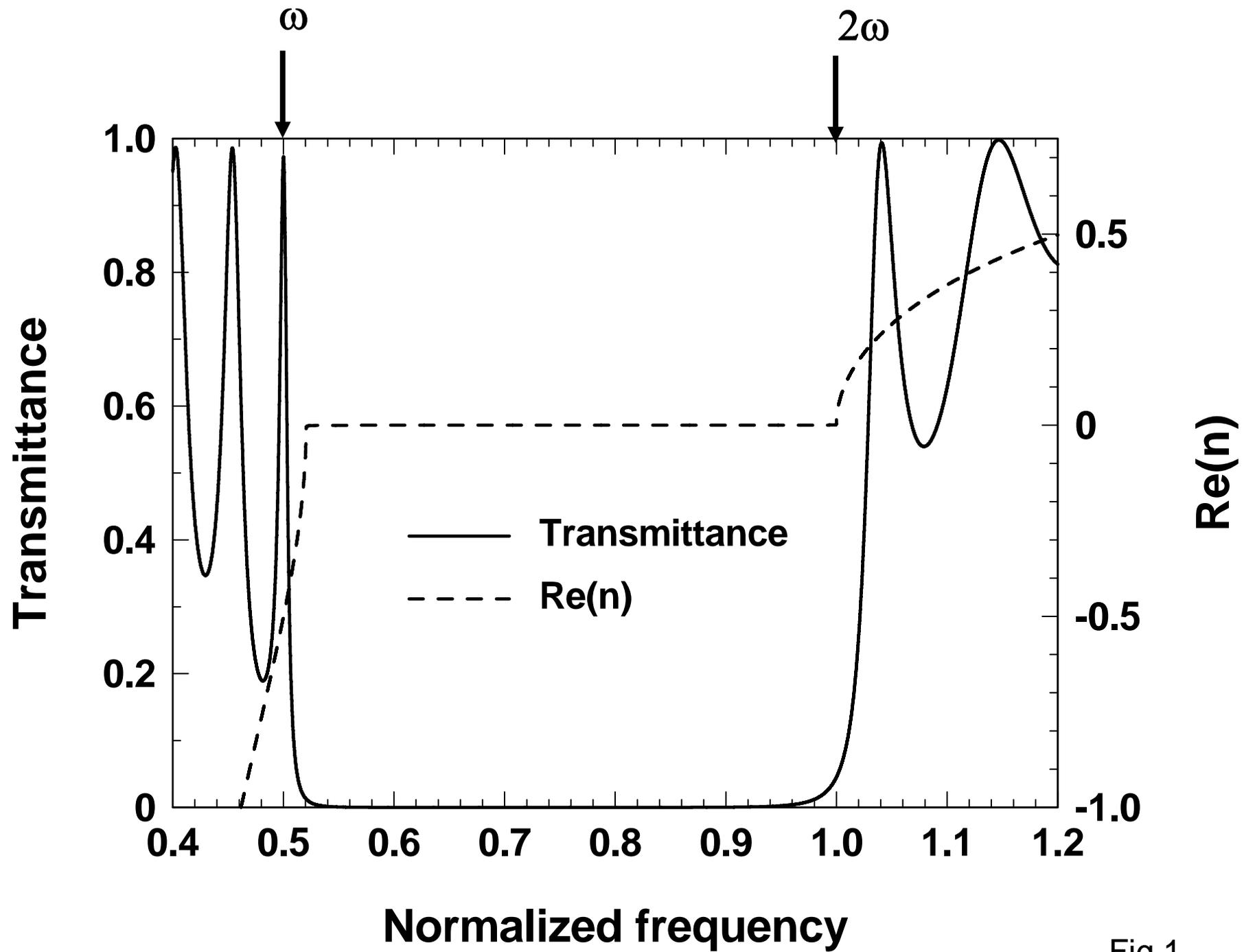

Fig.1

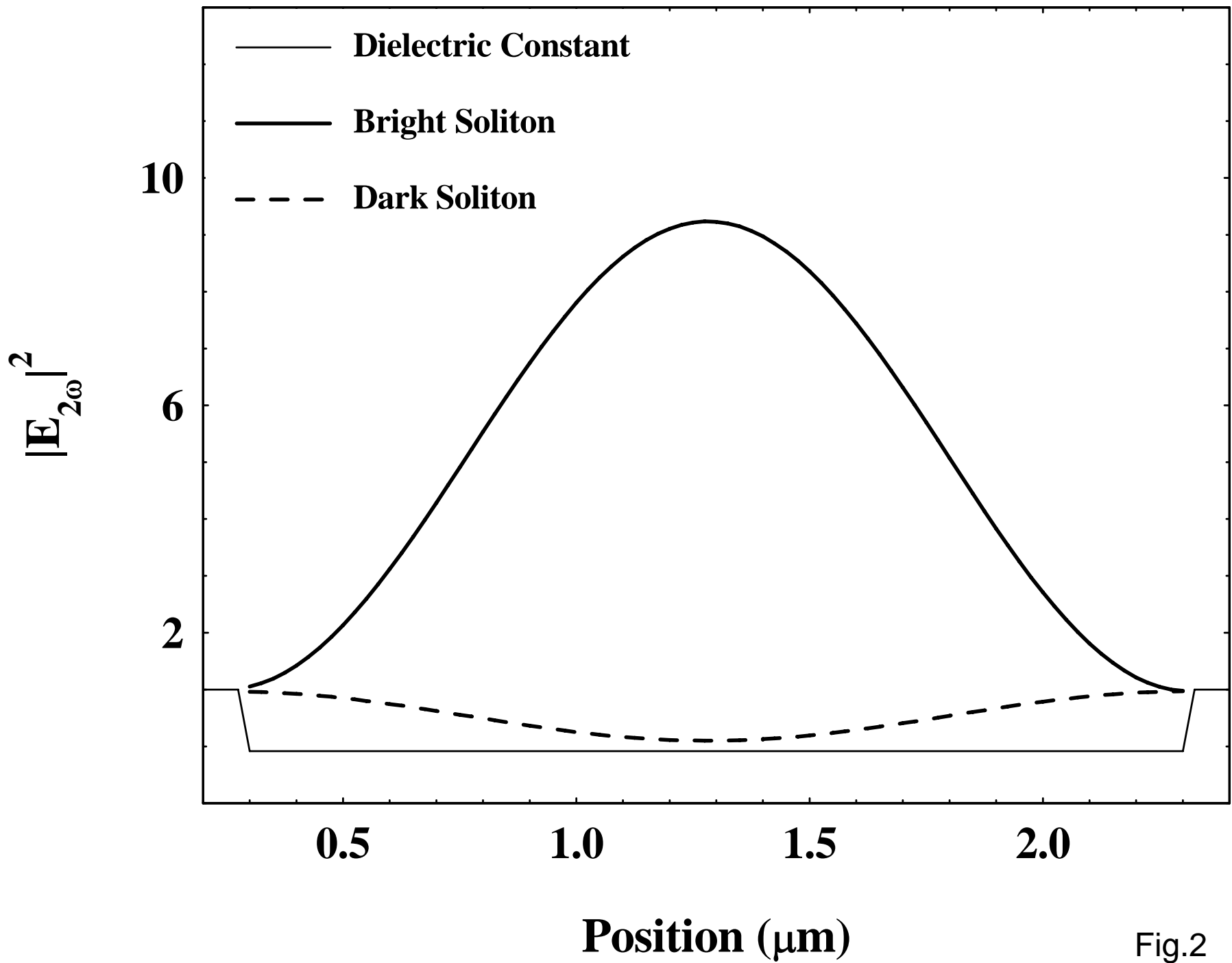

Fig.2

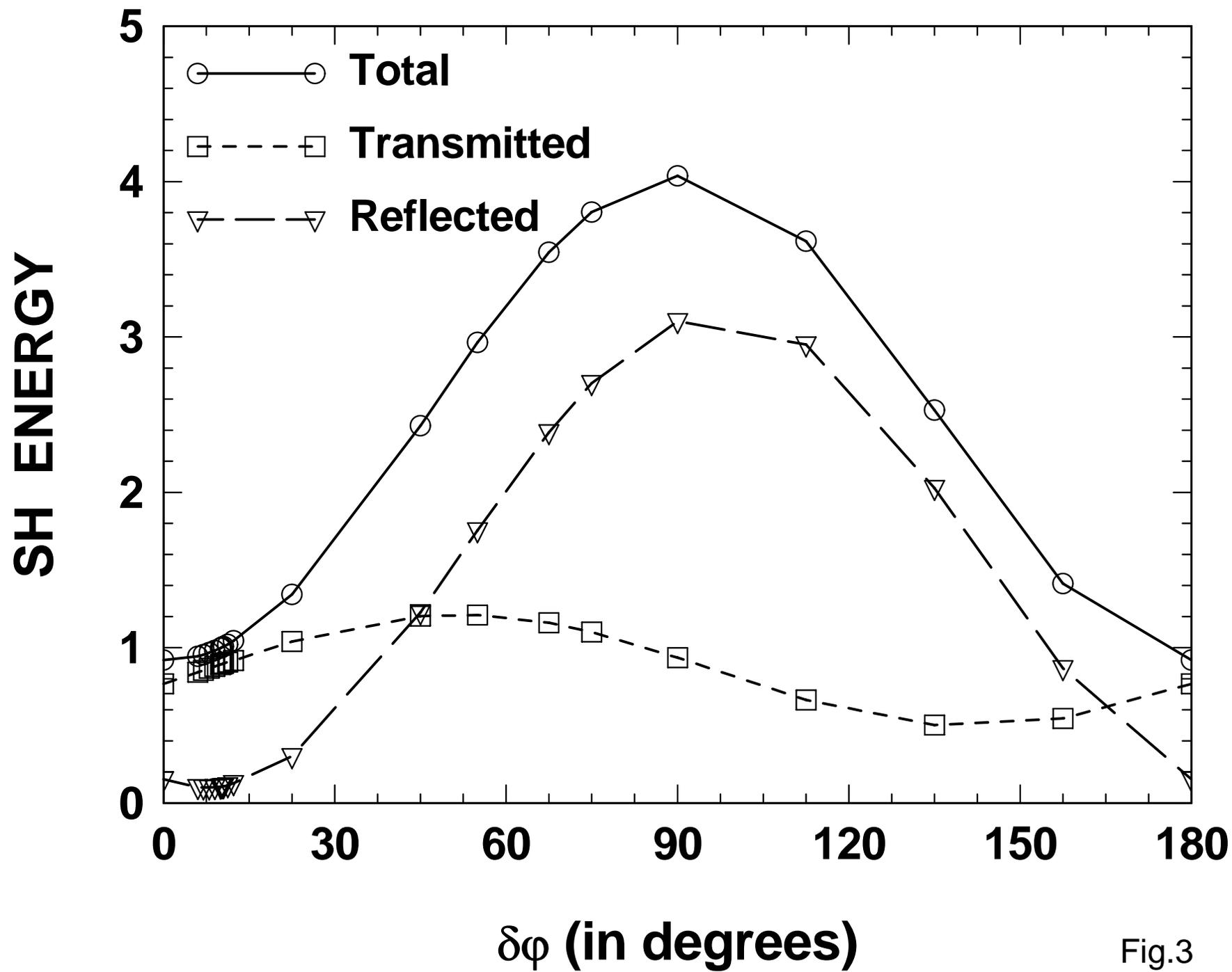

Fig.3

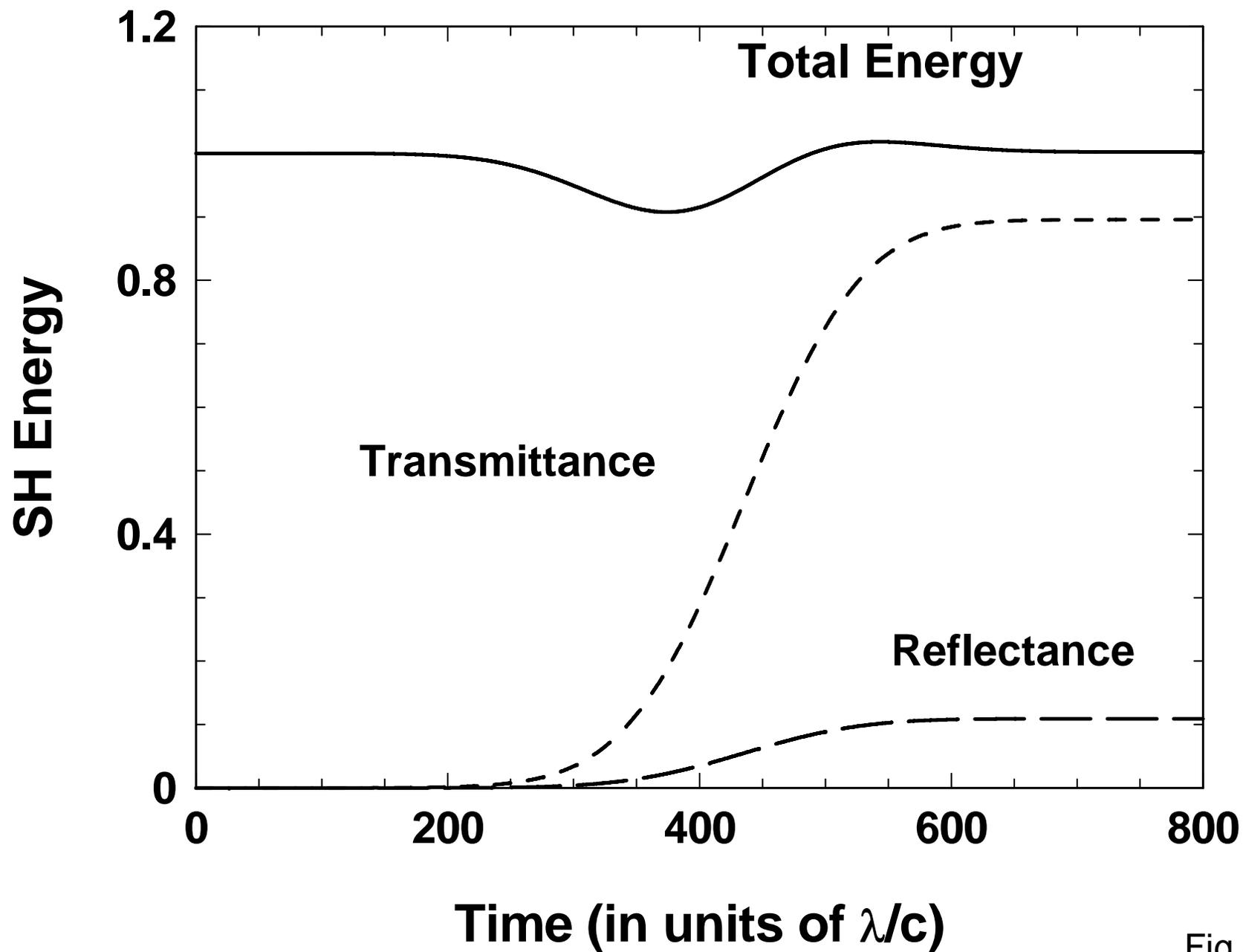

Fig.4

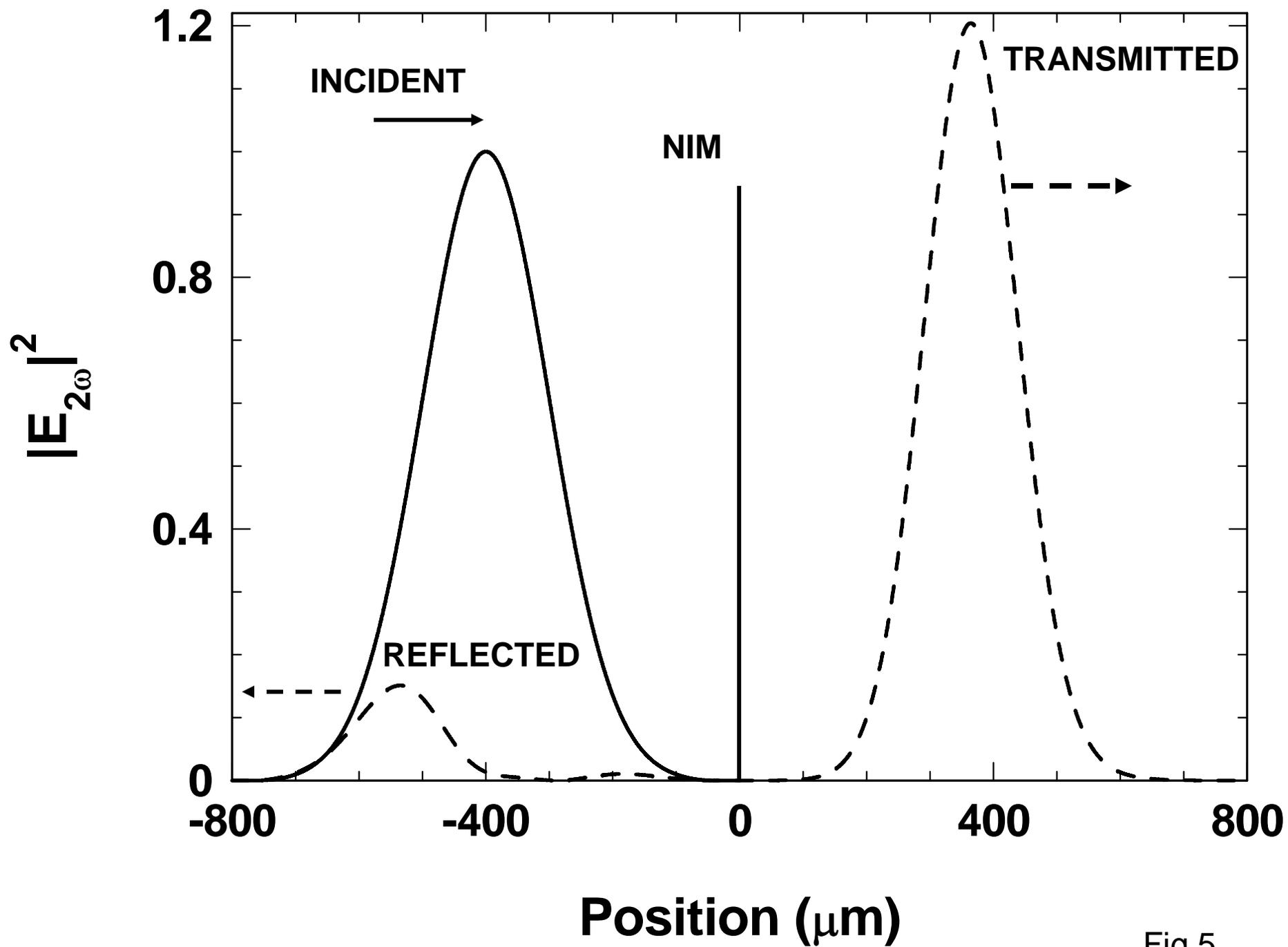

Fig.5